\documentstyle[11pt,epsfig]{article}
\title{\bf Horizon Problem Remediation via Deformed Phase Space}
\author{S.M. M. Rasouli\thanks{email: m-rasouli@sbu.ac.ir},\
        Mehrdad Farhoudi\thanks{email: m-farhoudi@sbu.ac.ir}\ \
        and Nima Khosravi\thanks{email: n-khosravi@sbu.ac.ir} \\
        {\small Department of Physics, Shahid Beheshti University, G.C.,
        Evin, Tehran 19839, Iran}}
\begin{document}
\date{\small April 26, 2011}
\maketitle
\begin{abstract}
We investigate the effects of a special kind of dynamical
deformation between the momenta of the scalar field of the
Brans--Dicke theory and the scale factor of the FRW metric. This
special choice of deformation includes linearly a deformation
parameter. We trace the deformation footprints in the cosmological
equations of motion when the BD coupling parameter goes to
infinity. One class of the solutions gives a constant scale factor
in the late time that confirms the previous result obtained via
another approach in the literature. This effect can be interpreted
as a quantum gravity footprint in the coarse grained explanation.
The another class of the solutions removes the big bang
singularity, and the accelerating expansion region has an infinite
temporal range which overcomes the horizon problem. After this
epoch, there is a graceful exiting by which the universe enters in
the radiation dominated era.
\end{abstract}
\medskip
{\small \noindent PACS number: $04.20.Fy$;
                               $04.50.Kd$; $02.40.Gh$; $98.80.Qc$}\newline
 {\small Keywords: Deformed Phase Space; Brans--Dicke Theory; Noncommutative Phase Space;
                   Quantum Cosmology.}
\bigskip
\section{Introduction}
\indent

General relativity~({\bf GR}) and quantum theory, as generally
believed, are two prominent paradigms to illustrate the nature.
The idea of GR was inspired by the general covariance, the
principle of equivalence and also the Mach ideas. However after
the formulation of GR established by Einstein, its full
satisfaction with the Mach principle has been a matter of debate.
That is, although the matter content of the universe affects the
geometry, but there are still vacuum solutions for GR, contrary to
the strong version of Mach idea that states if there is no matter
then, there will be no geometry. This proposition has made further
considerations for alternative gravitational theories to be more
Machian. One approach to this purpose has been stated by the
scalar--tensor theories which among them, the Brans--Dicke~({\bf
BD}) theory is the simplest one~\cite{BD61}. In this scope, a
scalar field plays the role of Newtonian gravitational constant
and makes the theory to be more Machian\rlap.\footnote{However,
the other motivation for development of this variant of GR was the
Dirac large number hypothesis~\cite{D38}.}\ Also in the BD theory,
there exists an adjustable dimensionless parameter which, in
principle, can be fixed by observations. On the other hand, with
suitable boundary conditions, GR is deducible from the BD theory
when making this parameter goes to infinity limit (though not
always, see, e.g., Refs.~\cite{BR93,BS97,Faraoni98,Faraoni99}). In
this view, the BD theory presents a modified version of GR (at
least for traceless energy--momentum tensor~\cite{Faraoni99}).
Hence in this sense, there may also exist some other features in
the BD theory in addition to those that are usually in GR. Such a
kind of features (if any) can be interpreted as internal
symmetries (or structures) of GR which are somehow missed if one
just starts purely with the Ricci scalar. This aspect can also be
viewed as if GR is only an effective theory, and the BD theory
when the $\omega$ goes to infinity is a more realistic one.

The quantum theory is mainly established to justify the behavior
of very small scales. Thus, it should be considered for small
scale behaviors of GR (which is an excellent theory for large
scale structures) as well. For example, in the standard cosmology,
the universe has been commenced by a big bang in the very early
universe. It is generally believed that, in the big bang, the
scale of the universe is almost zero, thus it is predictable that
the quantum behaviors to be significant in this regime. Indeed,
there have been considerable attempts to combine quantum theory
and GR in order to achieve a quantum gravity
theory~\cite{R.book04,A02,ASS04,SW99}. One of these approaches is
deformation of the phase space structure~\cite{ZachosFairlie05}
that introduces, if not all but at least, a part of the quantum
effects. This mechanism has been employed in the context of
cosmology~\cite{COR02,BP04}, in affecting the small and even large
scale behaviors, by removing the singularity and by the coarse
grained effects, respectively. In this approach, usually a
(length) parameter, which can be interpreted as the Planck
(length) constant, presents the quantum regime. However, it is
crucial to recover the standard results by taking the appropriate
limits of this parameter. The another significant aspect of this
kind of mechanism is its correspondence to the other methods of
quantization, which is~not only interesting but also
indispensable, see, e.g., Refs.~\cite{KHS09,KHSV10}.

As the scalar--tensor gravitational theories involve more degrees
of freedom, they can give more number of solutions than
GR~\cite{SST95}. Hence, we purpose to investigate deformation of
the phase space structure in this context by studying the flat FRW
metric and employing noncommutativity between the momenta of the
BD scalar field and the scale factor. Then, we obtain the BD
dynamical equations, however for simplicity and to be able to
compare the outcomes with the corresponding results in GR, we
solve them when the BD coupling parameter goes to infinity.
Though, it is important to note that such a procedure does~not
make a precise transition to GR when the trace of energy--momentum
vanishes~\cite{Faraoni99}, but we should also emphasize that if
one just starts the procedure by the GR formalism, one cannot
achieve such a wide classes of solutions, as expected.

In the next section, we briefly describe the BD theory in the
context of Hamiltonian formalism which is essential for
introducing deformation (noncommutativity) in the phase space. In
Section~$3$, we derive the BD equations of motion in the presence
of a special kind of deformation, then deduce their cosmological
implications when the BD coupling parameter goes to infinity.
Then, we discuss the solutions for different signs of the
deformation parameter and the integration constants of the
solutions while highlighting their effects. Finally, we will end
up the work by conclusions in the last section, while two
appendixes have also been furnished.

\section{Hamiltonian Formalism for Brans--Dicke Theory}
\indent

In this section, we review the BD theory in the Hamiltonian
formalism. However, in order to study the cosmological behavior,
we consider the spatially flat FRW metric as the background
geometry, namely
\begin{equation}\label{metric1}
ds^{2}=-N^2(t)dt^2+a^2(t)\left(dx^2+dy^2+dz^2\right),
\end{equation}
where $N(t)$ is a lapse function and $a(t)$ is the scale factor.

The BD Lagrangian density in the Jordan
frame~\cite{BD61,Jordan55,FGN99} in vacuum is given by
\begin{equation}\label{lag1}
{\cal L}=\sqrt{-g}\left(\phi R-
\frac{\omega}{\phi}g^{\mu\nu}\phi_{,\mu}\phi_{,\nu}\right),
\end{equation}
where the Greek indices run from zero to three. Also, the $\phi$
is the BD scalar field, $R$ is the Ricci scalar and $\omega$ is
the BD coupling constant that is supposed to be bigger than $-3/2$
for non--ghost scalar field
situations~\cite{BKM04,DDB07,BS07,B09}. Replacing the Ricci scalar
associated to metric (\ref{metric1}) into the above Lagrangian
gives
\begin{equation}\label{lag2}
{\cal L}=-6N^{-1}a\dot{a}^2\phi-6N^{-1} a^2\dot{a}\dot{\phi}+\omega
N^{-1}a^3\phi^{-1}\dot{\phi}^2,
\end{equation}
where the dot represents derivative with respect to the time and a
total time derivative term has been neglected. Thus, the
corresponding Hamiltonian is
\begin{equation}\label{Ham-1}
{\cal
H}_{0}=\frac{N}{\chi}\left(-\frac{\omega}{12}a^{-1}\phi^{-1}P_a^2+
\frac{1}{2}a^{-3}\phi P_\phi^2-\frac{1}{2}a^{-2}P_aP_\phi\right),
\end{equation}
where $\chi\equiv 2\omega+3$. As the momentum conjugate to $N(t)$
vanishes, one has to add it as a constraint to the above
Hamiltonian. Therefore, the Dirac Hamiltonian becomes
\begin{equation}\label{Dirac.H}
{\cal H}={\cal H}_{0}+\lambda P_N,
\end{equation}
where $\lambda$ is a Lagrange multiplier. Here, we consider the
ordinary phase space structure described by the usual ordinary
Poisson brackets
\begin{equation}\label{Poiss.eq1}
\{a,P_a\}=\{\phi,P_\phi\}=\{N,P_N\}=1,
\end{equation}
where the other brackets vanish. Therefore, the equations of
motion with respect to the Hamiltonian (\ref{Dirac.H}) are
\begin{eqnarray}\label{diff.eq1}
\dot{a}&=&\{a,{\cal
H}\}=-\frac{N}{\chi}\left(\frac{\omega}{6}a^{-1}\phi^{-1}P_a+
\frac{1}{2}a^{-2}P_\phi\right),\\
 \label{diff.eq2}
\dot{P_a}&=&\{P_a,{\cal
H}\}=-\frac{N}{\chi}\left(\frac{\omega}{12}a^{-2}\phi^{-1}P_a^2-
\frac{3}{2}a^{-4}\phi
P_\phi^2+a^{-3}P_aP_\phi\right),\\
 \label{diff.eq3}
\dot{\phi}&=&\{\phi,{\cal H}\}=\frac{N}{\chi}
\left(a^{-3}\phi P_\phi- \frac{1}{2}a^{-2}P_a\right),\\
 \label{diff.eq4}
\dot{P_\phi}&=&\{P_\phi,{\cal H}\}=-\frac{N}{\chi}
\left(\frac{\omega}{12}a^{-1}\phi^{-2}P_a^2+
\frac{1}{2}a^{-3}P_\phi^2\right), \\
 \label{diff.eq5}
\dot{N}&=&\{N,{\cal H}\}=\lambda, \\
 \label{diff.eq6}
\dot{P_N}&=&\{P_N,{\cal H}\}=\frac{1}{\chi}
\left(\frac{\omega}{12}a^{-1}\phi^{-1}P_a^2- \frac{1}{2}a^{-3}\phi
P_\phi^2+\frac{1}{2}a^{-2}P_a P_\phi\right).
\end{eqnarray}

Let us work in the comoving gauge, that is we fix the gauge by
$N=1$. Also, to satisfy the constraint $P_N =0$ at all times, the
secondary constraint $\dot{P_N}= 0$ should also be satisfied.
Hence, by Eq.~(\ref{diff.eq6}), one obtains
\begin{equation}\label{P.phi}
P_\phi=\frac{1}{2}\left(1\pm\sqrt{\frac{\chi}{3}}\right)a\phi^{-1}P_a.
\end{equation}
Now, differentiating Eq.~(\ref{diff.eq1}) with respect to the
time, while using Eqs.~(\ref{diff.eq2}), (\ref{diff.eq3}) and
(\ref{P.phi}), leads to
\begin{equation}\label{diff.a}
 \ddot{a}=-\left(\frac{2\chi}
 {\chi\pm\sqrt{3\chi}}\right)a^{-1}\dot{a}^2\, .
\end{equation}
In addition to the trivial static solution, one can get solutions
as $a(t)=C_1(t-t_{\rm ini.})^{q_{\pm}}$ where $C_1$ and the
initial time $t_{\rm ini.}$ are integration constants, and
$q_{\pm}$ is
\begin{equation}\label{w}
 q_{\pm}=\frac{2}{3\chi-1}\left[\frac{\chi-1}{2}\pm\sqrt
 {\frac{\chi}{3}}\right],
\end{equation}
when $\chi\neq 1/3$ (i.e. $\omega\neq -4/3$). In what follows, we
assume that $t_{\rm ini.}=0$, and the constant $C_1$ can be fixed
by the scale of the universe at an appropriate definite time.
Also, one can easily obtain
\begin{equation}\label{phi-dot-1}
\frac{\dot{\phi}}{\phi}=\mp\left(\frac{2\sqrt{3\chi}}
{\chi\pm\sqrt{3\chi}}\right) \frac{\dot{a}}{a}\,,
\end{equation}
with solutions $\phi(t)=C_2t^{s_{\pm}}$, where $C_2$ is an
integration constant and $s_{\pm}$ is given by
\begin{equation}\label{s-p}
 s_{\pm}=\frac{2(1\mp\sqrt{3\chi})}{3\chi-1}\,,
\end{equation}
when again $\chi\neq 1/3$. Through the Hamiltonian approach, we
have actually rederived the O'Hanlon and Tupper
solution~\cite{o'hanlon-tupper-72,FBOOK04}, as expected. This
solution has a big bang singularity when $t$ tends to zero. Also,
we should emphasize that the behavior of the scale factor and the
scalar field depends on the BD coupling parameter. In particular,
when $\omega$ goes to infinity, the non--trivial
solution\footnote{One could easily obtain the solution from
Eq.~(\ref{diff.a}) for this limit, however for the sake of
completeness, we have firstly derived explicit solutions for any
$\omega$ too.}\
 is~not the same as the corresponding result of GR (i.e. the Minkowski
space--time), for in this case one gets
\begin{equation}\label{omega limit}
a(t)=C_1t^{1/3}\qquad {\rm and}\qquad \phi={\rm constant},
\end{equation}
where the scale factor has a decelerated expanding behavior.

In the case $\chi=1/3$, the solutions for upper sign are
$a=C_3(t-t'_0)^{2/3}$\ and $\phi=C_4(t-t'_0)^{-1}$, and for lower
sign are $a=a_0\exp(t/t''_0)$\ and $\phi=\phi_0\exp(-3t/t''_0)$,
where $C_3$, $C_4$, $a_0$, $\phi_0$, $t'_0$ and $t''_0$ are
constants.

In the next section, we investigate a modified version of the
above formalism by introducing a noncommutative model.

\section{Deformed Phase Space Brans--Dicke Structure}
\indent

As mentioned in the introduction, deformation of the phase space
can present a sort of tracing quantum footprints in a given
model~\cite{ZachosFairlie05}. Of course, a general deformation
makes equations very difficult and even unsolvable. Hence, it is
customary to pick a proper choice which not only makes
calculations be possible, but also gives non--trivial results.
Indeed, by appealing to the simplicity principle (or the Occam's
razor), simplifications are usually performed in most toy models,
and even in real ones, before a consistent and complete theory is
deduced. Thus in this work, we just consider a dynamical
deformation between the conjugate momentum sector as
\begin{equation}\label{noncum.eq1}
\{P'_a,P'_\phi\}=l\phi'(t)
\end{equation}
and leave the other Poisson brackets among the primed parameters
(corresponding to those appeared in relation~(\ref{Poiss.eq1}))
unchanged. Hence, the Jacobi identity is still satisfied. In
general, noncommutativity between the momenta is a kind of
generalization of the usual noncommutativity between the spatial
coordinates~\cite{DjeSma04}. In addition, noncommutativity between
the momenta, in effective, has similarity with the behavior of a
charged particle in the presence of a magnetic field. In the scope
of gravitational theories, this kind of noncommutativity can be
interpreted as a gravitomagnetic field~\cite{MFinprogress}. Also,
one may find a clue in the string theory, especially in the flux
compactification, as mentioned in Ref.~\cite{kss}. The dynamical
behavior of deformation as time dependence has also been employed
in the literature, e.g. the $\kappa$--Minkowskian
spacetime~\cite{kappa,kappa1} and the generalized uncertainty
principle~\cite{gup1,gup2}, that are considered also in the
cosmological phase space~\cite{KHS08,babak}. In addition, the main
reason for the peculiarity of the chosen term in the right hand
side of (\ref{noncum.eq1}) is the dimensionality analysis that is
described in more details in the Appendix~A.

The minimally deformed (noncommutative) version of Hamiltonian
(\ref{Ham-1}) is achieved by replacing the unprimed variables with
the primed ones, namely
\begin{equation}\label{ham.2} {\cal
H}'_0=\frac{N'}{\chi} \left(-\frac{\omega}{12}a'^{-1}\phi'^{-1}
P_a'^2+ \frac{1}{2}a'^{-3}\phi' P_\phi'^2
-\frac{1}{2}a'^{-2}P_a'P_\phi'\right).
\end{equation}
However, it is more convenient to re--introduce the new variables by
applying the standard transformation~\cite{DjeSma04,CST01}
\begin{equation}\label{trick.eq1}
P'_\phi=P_\phi-la\phi,
\end{equation}
where the other unprimed variables are equivalent to their primed
counterparts. By considering the above transformation,
relation~(\ref{noncum.eq1}) is satisfied when the unprimed
variables satisfy the ordinary Poisson brackets. Then,
substituting $P'_\phi$ from (\ref{trick.eq1}) into Hamiltonian
(\ref{ham.2}) gives
\begin{equation}\label{s}
{\cal H}_0^{\rm nc}={\cal H}_0 +\frac{Nl}{\chi}\left(\frac{l}
{2}a^{-1}\phi^3-a^{-2}\phi^2P_{\phi} +\frac{1}{2}a^{-1}\phi
P_a\right),
\end{equation}
where we have also substituted ${\cal H}'_0$, as a function of the
primed variables, with ${\cal H}_0^{\rm nc}$, as a function of the
unprimed ones via the employed transformation. Once again, the
noncommutative Dirac Hamiltonian is
\begin{equation}\label{Dirac.Ham.eq2}
 {\cal H}^{\rm nc}={\cal H}_0^{\rm nc}+\lambda P_N.
\end{equation}
Therefore, the equations of motion become
\begin{eqnarray}\label{mot.eq1}
\dot{a}&=&\{a,{\cal H}^{\rm nc}\}=
\frac{N}{\chi}\left(-\frac{\omega}{6}a^{-1}\phi^{-1}P_a-
\frac{1}{2}a^{-2}P_\phi+\frac{l}{2}a^{-1}\phi\right),\\
\dot{P_a}&=&\{P_a,{\cal H}^{\rm nc}\}=\nonumber\\
 \label{mot.eq2}
&&-\frac{N}{\chi}\left(\frac{\omega}{12}a^{-2}\phi^{-1}P_a^2-
\frac{3}{2}a^{-4}\phi P_\phi^2+a^{-3} P_aP_\phi
-\frac{l^2}{2}a^{-2}\phi^3+2la^{-3}\phi^2P_\phi
-\frac{l}{2}a^{-2}\phi P_a\right),\\
 \label{mot.eq3}
\dot{\phi}&=&\{\phi,{\cal H}^{\rm nc}\}=
\frac{N}{\chi}\left(a^{-3}\phi P_\phi-
\frac{1}{2}a^{-2}P_a-la^{-2}\phi^2\right),\\
 \label{mot.eq4}
\dot{P_\phi}&=&\{P_\phi,{\cal H}^{\rm nc}\}=
-\frac{N}{\chi}\left(\frac{\omega}{12}a^{-1}\phi^{-2}P_a^2+
\frac{1}{2}a^{-3}P_\phi^2+ \frac{3}{2}l^2a^{-1}\phi^2-2la^{-2}\phi
P_\phi+\frac{l}{2}a^{-1}P_a\right), \\
 \label{mot.eq5}
\dot{N}&=&\{N,{\cal H}^{\rm nc}\}=\lambda\,, \\
\dot{P_N}&=&\{P_N,{\cal H}^{\rm nc}\}=\nonumber\\
 \label{mot.eq6}
&&+\frac{1}{\chi}\left(\frac{\omega}{12}a^{-1}\phi^{-1}P_a^2-
\frac{1}{2}a^{-3}\phi P_\phi^2+\frac{1}{2}a^{-2}P_a P_\phi
-\frac{l^2}{2}a^{-1}\phi^3+la^{-2}\phi^2P_\phi-\frac{l}{2}a^{-1}\phi
P_a\right).
\end{eqnarray}

In the comoving gauge, i.e. $N=1$, the secondary constraint
$\dot{P_N}=0$ gives
\begin{equation}\label{P.phi2}
P_\phi=\frac{a}{2}\left[2l\phi
+\left(1\pm\sqrt{\frac{\chi}{3}}\right) \phi^{-1}P_a \right]\, .
\end{equation}
Employing Eqs.~(\ref{mot.eq1})--(\ref{mot.eq4}) and
(\ref{P.phi2}), and performing a little manipulation lead to
\begin{equation}\label{a-dot}
\ddot{a}=-\left(\frac{2\chi}{\chi\pm\sqrt{3\chi}
}\right)a^{-1}\dot{a}^2\pm\frac{1}{\sqrt{12 \chi}}\, la^{-2}\phi\,
\dot{a}
\end{equation}
and again
\begin{equation}\label{phi-dot}
\frac{\dot{\phi}}{\phi}=\mp\left(\frac{2\sqrt{3\chi}}{\chi\pm\sqrt{3\chi}}\right)
\frac{\dot{a}}{a}.
\end{equation}
Solution of Eq.~(\ref{phi-dot}) can be in the form
\begin{equation}\label{phi-sol}
 \phi=\phi_0\, a^{\xi},
\end{equation}
where $\phi_0$ is a constant and $\xi$ is
\begin{equation}\label{xi}
  \xi\equiv\mp\frac{2\sqrt{3\chi}}{\chi\pm\sqrt{3\chi}}.
\end{equation}
Substituting (\ref{phi-sol}) into (\ref{a-dot}) yields
\begin{equation}\label{a-dot2}
 \ddot{a}=-\left(\frac{2\chi}{\chi\pm\sqrt{3\chi}
}\right)a^{-1}\dot{a}^2\pm\frac{\phi_0}{\sqrt{12 \chi}}\,
la^{\xi-2}\dot{a}.
\end{equation}
Note that, when the deformation parameter $l$ tends to zero, all
noncommutative equations reduce to their corresponding ones in the
previous section.

Now, as proposed, we are interested to investigate effects when
$\omega$ goes to infinity, though again it does~not mean that it
makes transition to the standard GR, as has been shown for the
commutative case in the previous section. However, as it is
obvious from Eq.~(\ref{a-dot2}), in order to obtain such effects,
it crucially depends on how the constant $\phi_0$ is, or can be,
related to the $\omega$. Actually for this purpose, the value of
constant $\phi_0$ (which represents different initial conditions)
proportional to $\sqrt{\chi}$ can be a reasonable one. A
particular motivation for it, however, is the new term in
Eq.~(\ref{a-dot2}) (in comparison to Eq.~(\ref{diff.a})) which
should~not vanish in the limit $\chi\rightarrow\infty$ on the one
hand. Namely, if it would vanish, one would~not be able to see any
new effects in comparison to the original undeformed theory in the
limit $\omega\rightarrow\infty$. On the other hand, the new term
also should~not become infinite. However, a consequence of this
choice is that when $\omega$ tends to infinity, then $\phi_0$ goes
to infinity as well. Though, this brings a technical problem, i.e.
it makes some ambiguities in the behavior of $\phi$--field in
solution (\ref{phi-sol}) when $\phi_0 \longrightarrow \infty$ and
$\xi \longrightarrow 0$. In this limit, $\phi$ is a time
independent (constant) variable which is infinity. To study the
inconvenience caused by this divergence of $\phi_0$, the
renormalization argument may assist in the following manner.

It is well--known that the procedure of renormalization occurs in
the quantum field theoretical level. However in our toy model,
this may indicate itself naively only in the deformation parameter
as the only presenter of quantum regime in this work. Hence, as
the first option, the deformation parameter $l\equiv l_{\rm bare}$
can be re--defined in an appropriate way that makes the transition
from Eq.~(\ref{a-dot2}) to Eq.~(\ref{a-dot3}) being possible. That
is, it can be re--defined as $\ell_{\rm
renormalized}\equiv\phi_0\, l_{\rm bare}/\sqrt{12 \chi}$ (see
below Eq.~(\ref{a-dot3})) with a finite $\phi_0$. Then, when $\chi
\longrightarrow \infty$, the $l_{bare}$ deformation parameter goes
to infinity such that the $\ell_{\rm renormalized}$ deformation
parameter becomes a finite constant. Also, there is an alternative
approach which is considered in the Appendix~B.

Therefore, by taking $\omega$ goes to infinity and choosing the
minus sign\footnote{This choice is~not restrictive, for in the
following we will consider different signs for the $\ell$.}\
 in Eq.~(\ref{a-dot2}), one gets
\begin{equation}\label{a-dot3}
\ddot{a}=-2a^{-1}\dot{a}^2-\ell a^{-2}\dot{a},
\end{equation}
where $\phi_0 l=\sqrt{12\chi}\ell$, which fixes the new parameter
$\ell$ with dimensionality $L^{-1}$.
Substituting
$a^2\ddot{a}=(a^2\dot{a}\dot)-2a\dot{a}^2$ into Eq.~(\ref{a-dot3})
gives
\begin{equation}\label{trick}
(a^2\dot{a}\dot)=-\ell\dot{a}\,,
\end{equation}
that yields $a^2\dot{a}=-\ell a+C$, where $C$ is an integration
constant with dimensionality $[C]=L^{-1}$. Then, one easily
obtains
\begin{equation}\label{a-dot4}
\frac{1}{2}a^2+\frac{C}{\ell}a+\frac{C^2}{\ell^2}\ln\Bigl
|a-\frac{C}{\ell}\Bigr|=\ell(-t+t_0),
\end{equation}
where $t_0$ is an integration constant too. Obviously, the above
equation is invariant under the transformation
$(\ell,C,t)\rightarrow(-\ell,-C,-t)$. This symmetry makes a
counterpart relevant between the solutions and, consequently
reduces the number of investigations for different cases by
half\rlap.\footnote{For example, the case $\ell<0$\ and $C<0$\ is
the counterpart of the another case $\ell>0$\ and $C>0$\ when
$t\rightarrow -t$.}\
 Thus, in the following categorization,
we consider the two probable options (the Case~I and Case~II) of
the logarithmic term in Eq.~(\ref{a-dot4}) only for interesting
cases of different signs of the $\ell$ and $C$, without probing
the counterpart solutions. Also, for the sake of completeness, we
explicitly investigate the solutions when $\ell$ tends to zero in
Case~III.

\subsection{Case~I: $a-\frac{C}{\ell}>0$}
\indent

As mentioned, we investigate this case for different signs of the
$\ell$\ and $C$.

\subsubsection{Case~Ia: Negative $\ell$ \& $C$}
\indent

For convenience, assume $\tilde{\ell}\equiv -\ell>0$\ and $b\equiv
-C>0$, thus Eq.~(\ref{a-dot4}) reads
\begin{equation}\label{a-dot5}
\frac{1}{2}a^2+\vartheta a+\vartheta^2\ln\left(
a-\vartheta\right)=\tilde{\ell}(t-t_0),
\end{equation}
where $\vartheta\equiv C/\ell=b/\tilde{\ell}>0$ with valid
domain\footnote{We have neglected the equality $a=\vartheta$, for
it makes $t$ becomes $-\infty$.}\
 $a>\vartheta$. This means that the initial value of the scale
factor cannot be zero and indeed, the universe has been started
with a non--vanishing size. Therefore, one may interpret that the
existence of a deformation parameter, as an indicator which
usually presents the quantum corrections to models, removes the
big bang singularity. Actually, this result is a common
expectation in the quantum cosmological models.

Then, by differentiating (\ref{a-dot5}), for $a>\vartheta$, we
obviously get
\begin{equation}\label{fir.der}
\dot{a}=\frac{\tilde{\ell}}{a^2}(a-\vartheta)>0
\end{equation}
and hence,
\begin{equation}\label{con.rel}
\ddot{a}=-\frac{\tilde{\ell}^2}{a^5}(a-\vartheta)(a-2\vartheta).
\end{equation}
Thus, the sign of $\ddot{a}$ depends on two different regions,
$\vartheta<a<2\vartheta$\ and $a>2\vartheta$, which we investigate
in the following.\\

\noindent\textbf{Region} $\vartheta<a<2\vartheta$\\

In this region, the expansion is an accelerated one in similar to
the inflationary phase. Though, it is~not exactly as the standard
inflationary phase, but it can solve the horizon problem as will
be discussed in the following. Usually, a successful candidate for
the standard inflation should satisfy two essential properties
among the other ones, namely the $60$ e--fold duration and a
graceful exit from this epoch. Our model naturally satisfies the
latter requirement, for the scale factor transits to the next
region, i.e. $a>2\vartheta$, where it decelerates. However, at
first glance, it looks that it does~not satisfy the former
condition as it has much less than $60$ e--fold duration. Indeed,
the number of e--fold definition, i.e. $N=a_{\rm final}/a_{\rm
initial}$, for our model is $N=2\vartheta/\vartheta=2$.
Nevertheless, its result is comparable to the standard inflation
one by presenting a solution for the horizon problem, which we
indicate it after a brief review on the successes of the standard
inflation while clarifying the horizon problem.

It is well--known that the most important problem of the standard
cosmology, which is solved by proposition of an inflationary
scenario, is the horizon problem. Of course, the inflation also
solves the relic particle abundances (or the monopoles) and the
flatness problems. However, it is generally believed that among
these problems, the horizon problem is the most important one, for
at least there are alternative scenarios that can resolve the
other two problems in the same manner as the inflation
does~\cite{M.book05,W.book08}. The horizon problem arises when the
universe is observed to be isotropic and homogeneous in the large
scale structure. This requires that the initial conditions must be
in a way which give such a universe. The problem with the standard
cosmology is that although the matter fluctuations have been
inhomogeneous at the initial level, but these fluctuations did~not
have enough time for interactions and transforming information
about their situations. Consequently, the inhomogeneous initial
conditions should naturally result in inhomogeneous present large
scale structure which is in contradiction to the observations. The
inflationary idea solves this problem by taking a homogeneous part
of the initial condition, and inflates it to an appropriate size
for the beginning of the radiation dominated era. In our model,
the horizon problem is solved in another way.

Actually, the accelerating phase occurs  during $a_{\rm
initial}\longrightarrow\vartheta$ and $a_{\rm final}=2\vartheta$,
which is from $t_{\rm initial}\longrightarrow-\infty$ to a finite
final time, that is $t_{\rm final}=\vartheta^2(4+\ln
\vartheta)/\tilde{\ell}$ (assuming $t_0=0$). This means that the
accelerating phase takes infinite time, $\bigtriangleup t=(t_{\rm
final}-t_{\rm initial})\longrightarrow\infty$, and during this
phase, the matter fluctuations can interact with the other parts
of initial conditions, exchange information about their local
structures, and hence, approach to an equilibrium state which is
presented by a homogeneous structure. Thus, although in our simple
model, the accelerating phase cannot be interpreted as the
standard inflationary era, but it can address the horizon problem
of the standard cosmology. This quasi static\footnote{For it takes
infinite time to double the value of the initial scale factor.}\
 accelerating phase is very similar
to the Hagedorn phase of string gas cosmology~\cite{NBV06}.\\

\noindent\textbf{Region} $a>2\vartheta$\\

In this case, the expansion of universe is decelerating, and when
$a\longrightarrow \infty$, the first term in Eq.~(\ref{a-dot5}) is
the dominant one, hence in this limit, $a(t)$ tends to $t^{1/2}$\
that behaves as the radiation era. This phase occurs exactly after
the above accelerating phase, and can be interpreted as the
radiation dominated phase after the usual inflationary epoch in
the standard cosmology. Indeed, this result is completely in
agreement with what is usually proposed for the universe in
different cosmological models, the standard cosmology with or
without an inflation.

The behavior of the scale factor for Case~Ia is plotted as a solid
line in Fig.~$1$.
\begin{figure}
\begin{center}
\epsfig{figure=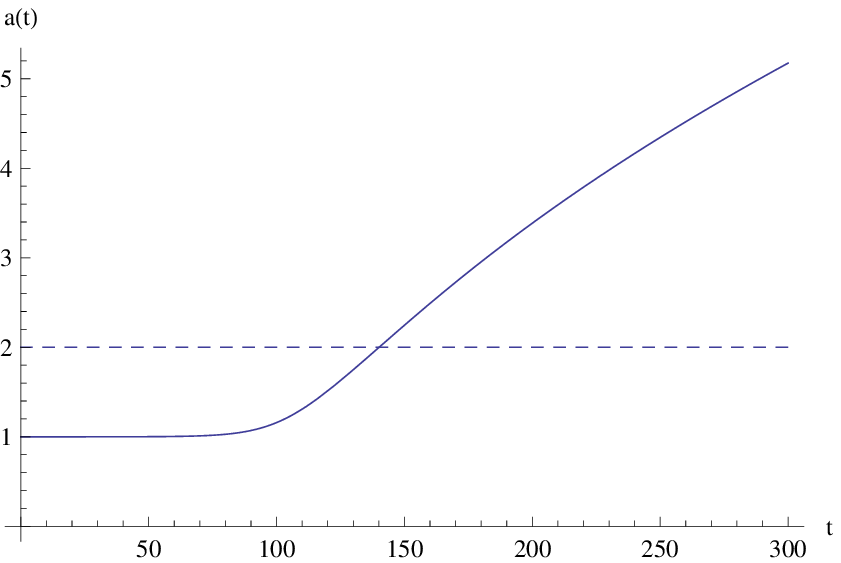,width=8cm}\hspace{5mm}
\epsfig{figure=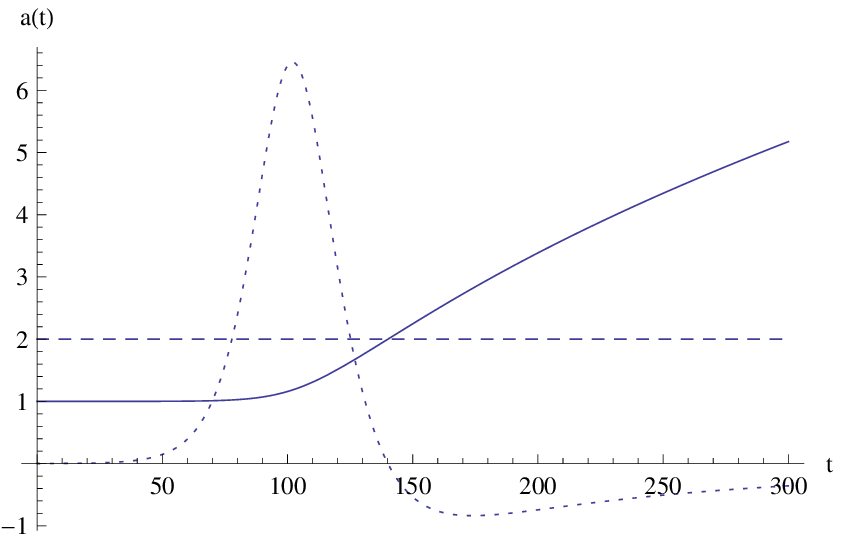,width=8cm}\hspace{5mm}

{\footnotesize \textbf{Fig. 1}: The solid line shows the behavior
of the scale factor for Case~Ia with $\ell=C=-0.1$. Below the
dashed line (i.e. $\vartheta<a<2\vartheta=2C/\ell=2$), one has an
accelerating phase, and above it (i.e. $a>2\vartheta=2C/\ell=2$),
a decelerating phase. In the right figure, the dotted curve
represents the $\ddot{a}(t)$ which is negative for
$a>2\vartheta=2$. Note that, the $\ddot{a}(t)$ curve has been
rescaled for a better clarification.}
\end{center}
\end{figure}

\subsubsection{Case~Ib: Negative $\ell$ \& Positive $C$}
\indent

This case is very similar to the previous one, except the sign
change in the acceleration. Hence, for the entire valid region of
the scale factor, i.e. $a>0$, the $\dot{a}>0$\ and $\ddot{a}<0$\
give a decelerated expanding behavior. When $t\rightarrow \infty$\
then $a(t)$\ tends to $t^{1/2}$, exactly as the previous case,
however here, there are~not interesting properties. The behavior
of the scale factor is depicted in Fig.~$2$(left).

\subsection{Case~II: $a-\frac{C}{\ell}<0$}
\indent

Once again, we investigate this case for different signs of the
$\ell$\ and $C$ too.

\subsubsection{Case~IIa: Positive $\ell$ \& $C$}
\indent

To describe this case, let us repeat Eq.~(\ref{a-dot4}) as
\begin{equation}\label{a-dot44}
\frac{1}{2}a^2+\vartheta
a+\vartheta^2\ln\left(\vartheta-a\right)=\ell(-t+t_0),
\end{equation}
where the valid domain of the scale factor is $0\leq a<\vartheta$,
which the $a\geq 0$ is a physical constraint. That is, in this
case, there is no regulator to prevent the big bang singularity,
for $a=0$ occurs at $t=t_0-(\vartheta^2\ln \vartheta)/\ell$.
However, the upper bound of the scale factor approaches to a
constant when $t$ goes to infinity, i.e. for the late time, one
gets
\begin{equation}\label{limit}
a\left(t\rightarrow\infty\right)=\vartheta={\rm constant}.
\end{equation}
This approaching to a constant value is a direct consequence of
the existence of a deformation parameter, and more interestingly,
it shows itself far from the big bang. As mentioned before, since
this deformation parameter may be interpreted as a consequence of
the quantum effects, then this feature may also be viewed as a
quantum gravity effect when the scale of the universe is
significant. That is, this behavior can be a phenomenological
property for the quantum gravity. However, these kind of
modifications can be perceived as a semi--classical model or, as a
model beyond the BD theory but still in the classical regime. The
scale factor decelerates in this choice, and its diagram is
plotted in Fig.~$2$(right). The graph shows that the behavior of
the scale factor is in agreement with the results obtained by
Ref.~\cite{KHS08} in where a dynamical deformation between the
lapse function and the scale factor has been employed.
\begin{figure}
\begin{center}
\epsfig{figure=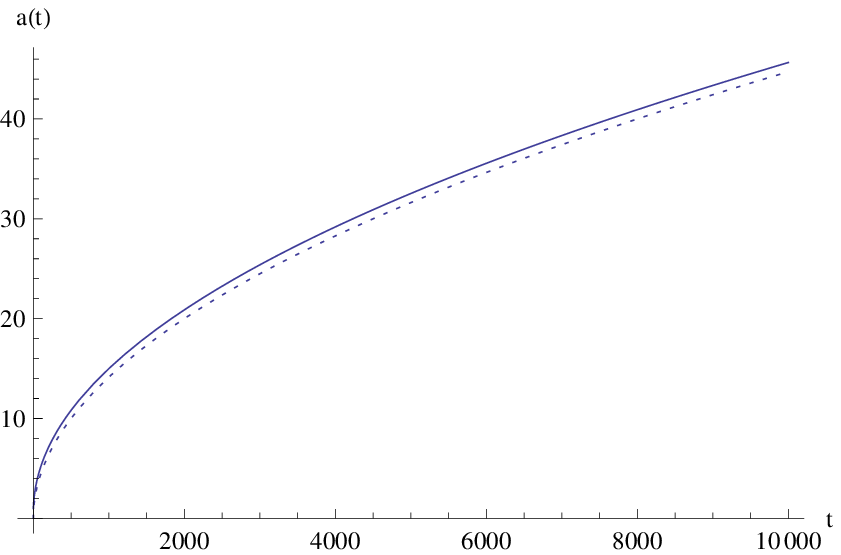,width=8cm}\hspace{5mm}
\epsfig{figure=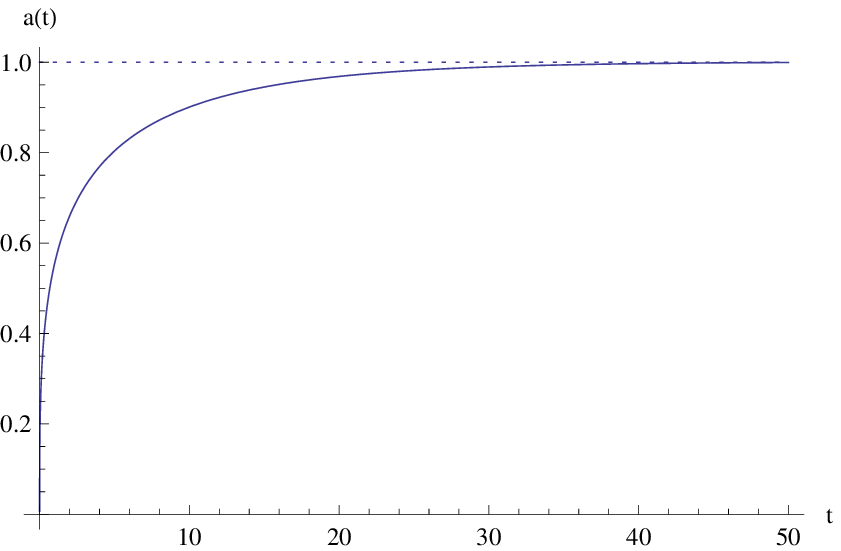,width=8cm}\hspace{5mm}

{\footnotesize \textbf{Fig. 2}: The solid line in the left figure
shows the behavior of the scale factor for Case~Ib with
$\ell=-C=-0.1$. As it is obvious, the scale factor asymptotically
is proportional to $t^{1/2}$ (the dotted line). The solid line in
the right figure shows the behavior of the scale factor for
Case~IIa with $\ell=C=0.1$. When $t\rightarrow\infty$, the scale
factor approaches $\vartheta=C/\ell=1$.}
\end{center}
\end{figure}

\subsubsection{Case~IIb: Positive $\ell$ \& Negative $C$}
\indent

The case $l>0$ and $C<0$ gives negative values for the scale
factor which is~not acceptable.

\subsection{Case~III: $\ell$ Tends to Zero}
\indent

As mentioned before, all equations reduce to their corresponding
commutative ones when the deformation parameter vanishes. Now, let
us explicitly investigate it for Eq.~(\ref{a-dot4}). Hence, taking
the limit $\ell\longrightarrow 0$ for the logarithmic term in
Eq.~(\ref{a-dot4}) gives
\begin{equation}\label{mac}
\frac{C^2}{\ell^2}\ln(\frac{C}{\ell}- a)=\frac{C^2}{\ell^2}\ln
\frac{C}{\ell}-\frac{Ca}{\ell}-\frac{a^2}{2}-\frac{\ell
a^3}{3C}-\cdots ,
\end{equation}
where higher order terms in $\ell$ can be neglected. And
obviously, the second and the third terms in the above relation
cancel the second and the first terms in Eq.~(\ref{a-dot4}),
respectively. The first term in relation (\ref{mac}) is a constant
and can be absorbed by re--definition of $t_0$. Consequently, the
scale factor tends to $t^{1/3}$ which recovers the commutative
solution (\ref{omega limit}) when $\omega$ goes to infinity, as
expected.

\section{Conclusions}
\indent

We have introduced a deformation in the phase space structure of
the two existing fields of the BD theory in the spatially flat FRW
metric. Also, we have traced the quantum footprints in the
cosmological equations of motion in the comoving gauge. All the
noncommutative equations are shown that do reduce to their
corresponding counterparts when the deformation parameter tends to
zero, as expected. Then as proposed, we have investigated the
effects when the BD coupling parameter goes to infinity. In this
process, we have faced an integration constant that depends on the
initial conditions, however in order to be able to trace the
effects, we assume it to be proportional to the square root of the
BD coupling parameter. We have also discussed our justifications
for why we have to fix it in this way and to render other side
effects.

Finally, different cosmological results have been deduced due to
the different possible signs for the two arbitrary parameters of
the solutions, namely the deformation parameter, $\ell$, and the
another integration constant (i.e. the $C$ in Eq.~(\ref{a-dot4})).
For one class of the solutions, the result predicts a constant
value for the scale factor in the late time that is in agreement
with the results obtained in Ref.~\cite{KHS08}. This feature may
be interpreted as a quantum gravity footprint in the large scale.

A more interesting result is achieved by the another class of the
solutions. In this case, it is shown that the existence of the
deformation parameter (or equivalently, the quantum correction)
removes the big bang singularity by preventing the scale factor
tends to zero. This also has an infinite temporal range for an
accelerating expansion region. However, this phase is~not the
standard inflationary phase, for its e--fold duration is a very
small number, but it can appropriately overcome the horizon
problem that is the main one in the standard cosmology. Indeed,
due to the very long time duration of this phase, the matter
fluctuations can transmit their information to the other parts of
the universe, and consequently become homogeneous. Implicitly,
after this epoch, there is a graceful exiting and then, the
universe enters in a radiation dominated era which is naturally in
agreement with the standard cosmology, with or without an
inflation. Note that, these consequences are held just by
introducing a constant parameter without considering any potential
in the model. It should also be mentioned that the model just
makes a (classical) background plausible to address the horizon
problem similar to (classical) background of the inflationary
models.

However, one of the major success of the standard inflation is its
prediction of (quantum) fluctuations' behavior. The standard
inflation anticipates a scale invariant spectral index which is in
good agreement with the observations. For our model, considering
the fluctuation of dynamics is important, for not only to compare
with the observational data but also, to test the stability of the
model. It should be checked whether inhomogeneous arbitrary
initial conditions can have a significant effect in the late time
behavior or not. To overcome such general questions, one needs to
perform more investigations, perhaps employing the perturbative
analysis which in our model is still more complicated than in the
standard inflation, due to the existence of the BD scalar field as
well as the deformation parameter. These are interesting
investigations for further considerations, and are~not in the
scope of the current work.

In fact the above achievement in solving the horizon problem can
be viewed as a natural consequence of the noncommutativity
approaches. That is, it is well--known that the varying speed of
light and noncommutative models are related to each
other~\cite{vsl}, where the first motivation for the former models
has been raised in order to achieve an alternative approach to the
standard inflation. Indeed, the coordinate noncommutativity has
been employed in the cosmological context for the same purpose as
well, see, e.g., Refs.~\cite{brandenberger,brandenberger1}.

\section*{Appendix A: On Dynamical Deformation As Relation
          (\ref{noncum.eq1})}
\indent

Let us first indicate the dimension of the Poisson bracket
$\{P_a,P_\phi\}$. In this work, we have employed the units
$\hbar=1=c$, therefore, from the Plank length, $l_P=\sqrt{\hbar
G/c^3}$, the dimensions of $G$ and $\phi$ are $[G]=L^2$ and
$[\phi]=L^{-2}$. The scale factor and the lapse function are
dimensionless parameters, the dimensions of coordinates and the BD
Lagrangian are $[x^\mu]=L$ and $[{\cal L}]=L^{-4}$. Hence, one can
conclude that $[P_a]=L^{-3}$ and $[P_\phi]=L^{-1}$, and
consequently $[\{P_a,P_\phi\}]=L^{-1}$. On the other hand, the
dimension of the deformation parameter is $[l]=L$. Therefore, from
dimensionality aspects of view, the dynamical deformation
(\ref{noncum.eq1}) is a plausible choice. Besides, from simplicity
point of view, with this choice, no other extra field has been
introduced in the model.

Of course, one may also propose other choices that still can
satisfy the dimensionality of $\{P_a,P_\phi\}$, but the suggested
relation (\ref{noncum.eq1}) is a first order (linear) term in the
deformation parameter as well. This suggestion is also a length
indicator that can present and trace the quantum behaviors, and if
the length indicator vanishes, one will recover the standard
(classical) counterpart relations. On the other hand, in
cosmological models a length parameter (e.g. the Planck length
that is a function of $\hbar$) is physically a more realistic and
plausible choice. That is, among different choices that can be
selected as a quantum indicator, a length scale is an appropriate
one for cosmological models, that can compare different scales for
the quantum or classical aspects as well.
\setcounter{equation}{0}
\renewcommand{\theequation}{B.\arabic{equation}}
\section*{Appendix B: Discussion on Fixing Integration Constant $\phi_0$}
\indent

To fix the inconvenience behavior of the $\phi_0$ when
$\chi\rightarrow\infty$, one may also apply the renormalization
procedure alternatively to the matter field $L_{\rm matter}$.
However in our model, there is no matter field and $G$ (which is
equal to $1/\phi$ in the BD theory) does~not appear in the
equations of motion and makes its value non--effective, but the
above procedure can be applied as if the matter field is turned
on. Hence, in this case, the value of Newtonian gravitational
constant can be regularized by re--definition of the matter field
as well as the BD scalar field. To be more specific, in the
presence of a matter field, the BD Lagrangian is
\begin{equation}\label{lag1matter}
{\cal L}=\sqrt{-g}\left(\phi R-
\frac{\omega}{\phi}g^{\mu\nu}\phi_{,\mu}\phi_{,\nu}+L^{\rm
bare}_{\rm matter}\right).
\end{equation}
Now, as an overall constant has no role in the form of equations
of motion, one can multiply the above BD Lagrangian by a
dimensionless parameter $(\phi_0 G_{bare})^{-1}$ to get
\begin{eqnarray}\label{lag1matter1}
 &&\frac{1}{\phi_0 G_{\rm bare}}\sqrt{-g}\left(\phi R-
   \frac{\omega}{\phi}g^{\mu\nu}\phi_{,\mu}\phi_{,\nu}+L^{\rm
   bare}_{\rm matter}\right)=\cr
 &&\sqrt{-g}\left[\frac{\phi}{\phi_0
   G_{\rm bare}} R- \omega\left({\frac{\phi}{\phi_0
   G_{\rm bare}}}\right)^{-1}g^{\mu\nu}\left(\frac{\phi}{\phi_0G_{\rm bare}}\right)_{,\mu}
   \left(\frac{\phi}{\phi_0G_{\rm bare}}\right)_{,\nu}+\frac{1}{\phi_0G_{\rm bare}}L^{\rm
   bare}_{\rm matter}\right].
\end{eqnarray}
The $\phi_0$, that goes to infinity when $\omega \longrightarrow
\infty$, can be absorbed in re--definition of $\phi$ and $L^{\rm
bare}_{\rm matter}$ by a renormalization process such that
\begin{eqnarray}\label{lag1matter11}
{\cal L}^{\rm renormalized}=\sqrt{-g}\left[\bar{\phi} R-
\frac{\omega}{\bar{\phi}}g^{\mu\nu}\bar{\phi}_{,\mu}\bar{\phi}_{,\nu}+L^{\rm
renormalized}_{\rm matter}\right],
\end{eqnarray}
where $\bar{\phi}= G^{-1}_{\rm renormalized}=\phi_{\rm
renormalized}\equiv\phi/(\phi_0G_{\rm bare})$ and $L^{\rm
renormalized}_{\rm matter}\equiv L^{\rm bare}_{\rm
matter}/(\phi_0G_{\rm bare})$. Thus, the difficulty of infinite
value of $\phi_0$ can be solved, at least naively, by the
renormalization procedure.

\section*{Acknowledgement}
\indent

We would like to thank H. Firouzjahi and M.M. Sheikh--Jabbari for
fruitful discussions.

\end{document}